\documentclass[aps,prl,nobalancelastpage,superscriptaddress,twocolumn,longbibliography,nobibnotes,nofootinbib]{revtex4-1}

\usepackage[utf8]{inputenc}
\usepackage[american,british]{babel}
\usepackage[T1]{fontenc}
\usepackage[pdftex]{graphicx}  
\usepackage{xcolor}
\usepackage{dcolumn}
\usepackage{physics}
\usepackage{braket}
\usepackage{bm}
\usepackage{amsmath,amsthm,amssymb}
\usepackage{color}
\usepackage{verbatim}
\usepackage[normalem]{ulem}
\usepackage{dsfont}
\usepackage{hyperref}
\hypersetup{
 colorlinks=true,
 linkcolor=blue,
 anchorcolor = blue,
 citecolor = blue,
 filecolor = blue,
 urlcolor = blue
}

\def \be {\begin{equation}} 
\def \ee {\end{equation}} 
\def \l {\left(} 
\def \r {\right)} 
\def \la {\langle} 
\def \ra {\rangle}  

\date{}

\newcommand{\saclay}{Universit\'e Paris-Saclay, CNRS, LPTMS, 91405, Orsay, France.}

\begin{document}

\title{
Phase Transitions without gap closing in monitored quantum mean-field systems
}
\date{\today}

\author{Luca Capizzi} 
\affiliation{\saclay}

\author{Riccardo Travaglino} 
\affiliation{SISSA and INFN Sezione di Trieste, via Bonomea 265, 34136 Trieste, Italy}

\begin{abstract}

We investigate the monitored dynamics of many-body quantum systems in which projective measurements of extensive operators are alternated with unitary evolution. Focusing on mean-field models characterized by all-to-all interactions, we develop a general framework that captures the thermodynamic limit, where a semiclassical description naturally emerges. Remarkably, we uncover novel stationary states, distinct from the conventional infinite-temperature state, that arise upon taking the infinite-volume limit. Counterintuitively, this phenomenon is not linked to the closing of the Lindbladian gap in that limit. We provide analytical explanation for this unexpected behavior.

\end{abstract}

\maketitle 

\paragraph{Introduction ---}

The interplay between unitary dynamics and measurements in quantum many-body systems gives rise to a variety of exotic phenomena that have been under investigation during the last decade. For example, the measurement-induced phase transition (MIPT), characterized by a change of the qualitative behaviour of the entanglement in the quantum trajectories, has been characterized in random unitary circuits \cite{Li-2018,lcf-19,Skinner_2019,Zabalo-20,Zabalo-22,st-22}, free systems \cite{Biella2021,Turkeshi-21,Fava-23,Poboiko-23,Fava-24}, critical theories \cite{Rossini-vicari,murciano23} and long-range systems \cite{Minato-22,Muller-22,Block-22,Giachetti-23,Li-25,slcg-25}. Related phenomena are Dissipative Quantum Phase Transitions (DQPT) \cite{Mitra-06,Prosen-08,Diehl-10,Kessler-12,Carmichael-15,Rodriguez-17,Fitzpatrick-17,Fink-17}, where the properties of stationary states of open systems change abruptly as a function of the microscopic parameters. These are usually detected by the closing of the \textit{Lindbladian gap} in analogy with usual Quantum Phase Transitions of closed systems \cite{Sachdev-99}.

On the other hand, it has been recently understood that the spectral gap of many-body systems is not necessarily related to the typical relaxation time \cite{Mori-20, Marche-25}: this happens, for example, when non-hermitian skin effect occurs \cite{Yao-18,Haga-21,Gohsrich-25}. As a consequence, it is hard to tell whether, in general, the behaviour of the spectral gap can detect those transitions (as pointed out in Ref. \cite{Matsumoto-20}). While a few different novel probes of DQPT (and typical relaxation times) associated with the Lindbladian spectrum have been recently proposed  \cite{Mori-23,Haga-24}, to the best of our knowledge, there is no universal consensus.

In this work, we study the dynamics of mean-field models where projective measurements are alternated with unitary dynamics. We find that additional stationary states can appear in the thermodynamic limit; yet, this mechanism is not immediately captured by the finite-size Lindbladian spectrum, which indeed remains gapped. This finding strongly challenges the usual view on DQPT, suggesting that novel probes associated directly with the thermodynamic limit of the system are probably necessary to detect these phase transitions.

\begin{figure}[t]
 \includegraphics[width=\columnwidth]{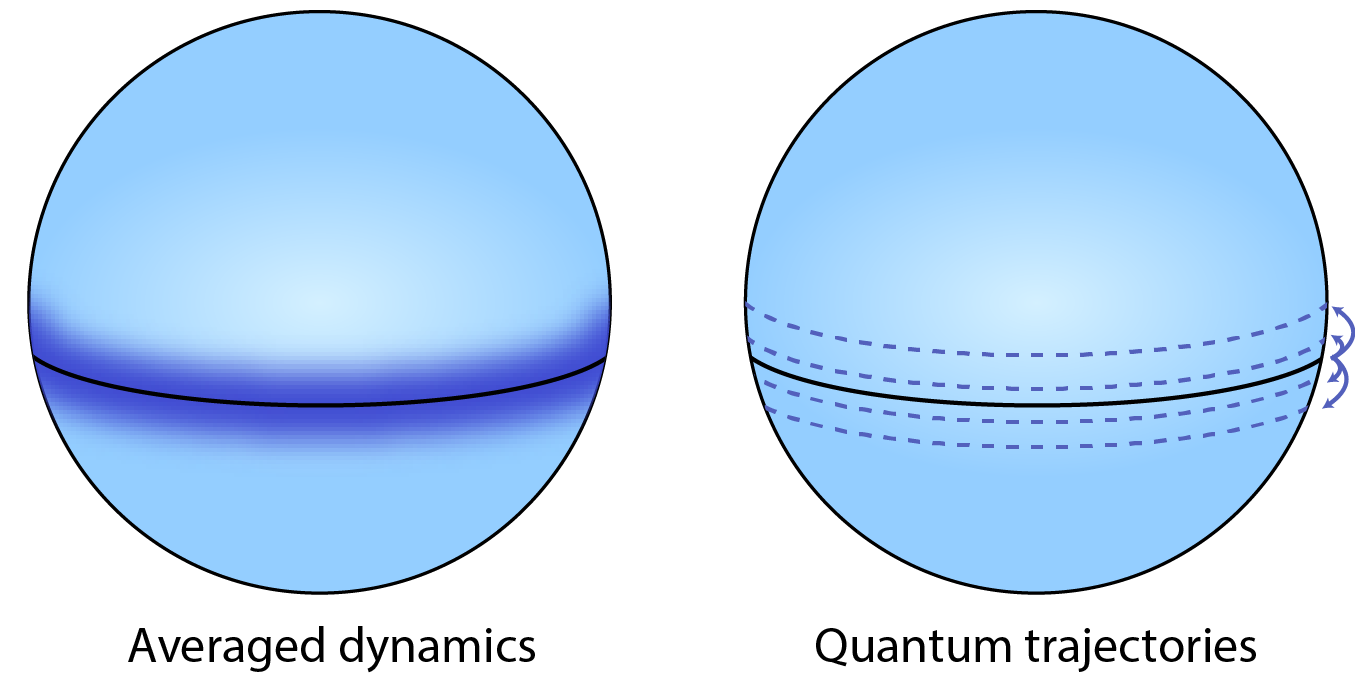}
 \caption{Monitoring via projective measurements of the total magnetization. Left (Averaged dynamics): A probability distribution with an initial given magnetization density $m^z$ spreads diffusively across the Bloch sphere. Right (Quantum trajectories): Quantum trajectories correspond to random walks of $m^z$; in any single realization, the state is always localized on the Bloch-sphere circumferences corresponding to the given $m^z$.
}
\label{fig:sketch}
\end{figure}
Specifically, we focus on projective measurements of extensive operators, a framework that is less explored in the context of MIPT, where local measurements are usually considered (although some results on global charges of free fermions \cite{Travaglino-25} and conformal field theories \cite{Rajabpour-15,Rajabpour-16,Lin-23,kv-25,mm-25} are available). We consider mean-field Hamiltonians, characterized by all-to-all interactions; these are interesting per se, due to their relation with long-range systems and the implementation of the corresponding experimental setups \cite{Defenu-23,Defenu-24,Monroe-21,Ritsch-13,Lahaye-09,Bohn-17,Weimer-10,Ferioli-23}, but they also provide the natural theoretical framework to access the thermodynamic limit \cite{lmg-95,Caneva-08,Ribeiro-08,Campbell-16,Russomanno-17}. In this context, we are able to study both the averaged dynamics and those of the quantum trajectories, in terms of an emerging Langevin dynamics (different from the monitoring with an external ancilla, studied in Refs. \cite{Fava-23,Fava-24,Li-25,slcg-25}), a sketch of which is given in Fig. \ref{fig:sketch}. 

\paragraph{The protocol ---}

We consider a spin-1/2 chain of length $L$. We measure an extensive operator $Q$ of the form
\be
Q = \sum_x q(x),
\ee
with $q(x)$ a strictly local operator supported on the site $x$ of the chain: we focus on $Q = S^z$, that is the total magnetization along the $z$-axis. The unitary dynamics is generated by the paradigmatic Ising Hamiltonian with longitudinal field
\be\label{eq:Ising}
H = \frac{J}{L}(S^x)^2+\lambda S^x;
\ee
here $J,\lambda$ denote the coupling and the field, respectively; the $1/L$ term in the Hamiltonian, also known as Kac's rescaling \cite{Kac-63}, ensures its extensivity.

Given $\rho$ a state, a measurement of $Q$ conditioned to the outcome $q$ is described by the projected state
\be
\rho \mapsto \frac{\Pi_q \rho \Pi_q}{\text{Tr}\l \rho \Pi_q\r},
\ee
with $q$ distributed according to the probability distribution $\{\text{Tr}\l \rho \Pi_q\r\}$ and $\Pi_q$ the associated projectors. After averaging over the measurement outcomes, we recover the \textit{averaged state}
\be\label{eq:symmetriz}
\rho \mapsto \sum_q \Pi_q \rho \Pi_q = \int^{2\pi}_0 \frac{d\alpha}{2\pi}e^{i\alpha Q}\rho e^{-i\alpha Q}.
\ee
We consider the unitary dynamics associated with a small time step $\tau$, described by the map $\rho \rightarrow e^{-iH\tau}\rho e^{i H \tau}$; the subsequent alternation of measurements of $Q$ with unitary evolution defines our protocol. While a single measurement can abruptly modify the state—resulting in a post-measurement state that is symmetric under $S^z$—the subsequent small unitary step produces a state that remains close to a symmetric one. Therefore, one could easily anticipate that a description in continuous time is possible at the level of symmetric states: we give the details below.

Let us consider a symmetric state $\rho$ ($[\rho,S^z]=0$); we first apply unitary dynamics for a time step $\tau$, expanding the generator at second order in $\tau$, and we symmetrize it (through Eq. \eqref{eq:symmetriz}). The first order gives unitary dynamics, on \textit{fast time scales}, associated with the symmetrized Hamiltonian $H_{\rm sym}=\sum_q \Pi_qH\Pi_q$: in our context, this acts trivially as the dynamics is constrained to the symmetric subspace. The second order is non-trivial and gives a Lindbladian dynamics; defining $\tau = \sqrt{dt}$  we obtain 
\be\label{eq:Lindblad}
\frac{d\rho}{dt} = - \frac{1}{2}\int^{2\pi}_0 \frac{d\alpha}{2\pi}[H_\alpha,[H_\alpha,\rho]]
\ee
restricted to the space of symmetric states, $H_\alpha := e^{i\alpha Q}He^{-i \alpha Q}$ being the rotated Hamiltonian. In the Supplemental Material, we will also comment on a slightly different protocol, in which the measurements are performed at a certain rate, showing analogies with \eqref{eq:Lindblad} when the measurement rate is large.

We anticipate that at $\lambda=0$ a phase-transition occurs: the initial states with spins aligned along the $z$-axis become stationary in the thermodynamic limit, giving rise to a finite-time relaxation time that grows with $L$. This mechanism is neither a consequence of symmetries nor easily detectable from the finite-size Lindbladian spectrum. To proceed, in the next sections we firstly develop a theoretical framework to study systematically the thermodynamic limit of these mean-field models.

\paragraph{Mean-field models and semiclassical limit---}

To provide a meaningful notion of the thermodynamic limit of the dynamics, we adopt an algebraic approach (standard in the mathematical formulation of infinite lattices \cite{rg-89} and mean-field models \cite{ola-87}). We focus on the algebra of observables $\mathcal{A}$ generated by the magnetization density\footnote{We refer to the $C^*$-algebra generated by finite sums/products of $\mathbf{m}$ with a topological closure induced by the operator norm; the involutive operation, necessary for the $*$-structure, is simply the hermitian conjugation $^\dagger$.}
\be
\mathbf{m} := \frac{\mathbf{S}}{L/2}, \quad \mathbf{S} = (S^x,S^y,S^z).
\ee
This algebra is commutative in the thermodynamic limit, since $[m^a,m^b]\sim 1/L\rightarrow 0$ for $L\rightarrow \infty$, and it is represented by the continuous functions from the unit ball $|\mathbf{m}|\leq 1$ to $\mathbb{C}$. A state $\la \dots\ra$, which is in general a positive functional $\mathcal{A}\rightarrow \mathbb{C}$, in this context is just a probability distribution over the ball. 
On the other hand, although the algebra $\mathcal{A}$ becomes commutative, the action of the Hamiltonian on $\mathcal{A}$ has a finite limit; namely, given
\be
H = \frac{L}{2}h(\mathbf{m}),
\ee
with $h$ a smooth function of $\mathbf{m}$, it is not difficult to show that, for any $\mathcal{O} \in \mathcal{A}$
\be
i[H, \mathcal{O}] = -\{h, \mathcal{O}\}_{\text{Poisson}}
\ee
where the Poisson structure on $\mathcal{A}$ is defined by
\be
\quad \{m^a,m^b\}_{\text{Poisson}} := \varepsilon^{abc} m^c,
\ee
with $\varepsilon^{abc}$ the Levi-Civita tensor (the summation over $c$ is implicit). From now on, we will omit the pedix in the Poisson bracket, denoting them by $\{\cdot,\cdot\}$.

This framework, which formalizes the emergence of the semiclassical limit for quantum mean-field models, applies specifically also to the Ising model in Eq. \eqref{eq:Ising} through the identification $h(\mathbf{m}) = J/2(m^x)^2+\lambda(m^x)$; for instance, the unitary dynamics associated with the Hamiltonian $\eqref{eq:Ising}$ gives rise to the classical equations of motion $\frac{d}{dt}\mathcal{O} = -\{h,\mathcal{O}\}$. Similar techniques can be employed to study the thermodynamic limit of \eqref{eq:Lindblad}. We represent a state $\la \dots \ra$ in terms of its probability density $\rho$, that is a function over the ball (and it can be though as an observable in $\mathcal{A}$), with respect to a reference invariant state $\la \dots \ra_0$ as follows
\be
\la \dots \ra := \frac{\la \rho \dots\ra_0}{\la \rho\ra_0}.
\ee
With this prescription, the dynamics of an expectation value of  $\mathcal{O}$ is given by $\la \mathcal{O}(t)\ra := \la \rho(t)\mathcal{O}\ra_0/\la \rho\ra_0$ where $\rho(t)$ evolves according to
\be\label{eq:lind_classical}
\frac{d\rho}{dt} = \frac{1}{2}\int^{2\pi}_0 \frac{d\alpha}{2\pi}\{h_\alpha,\{h_\alpha,\rho\}\},
\ee
describing the classical counterpart of Eq. \eqref{eq:Lindblad}. Here $h_\alpha$, the rotated hamiltonian density, is defined from the flow $\frac{d}{d\alpha}h_\alpha = -\{m^z,h_\alpha\}$.

We now describe the thermodynamic limit of our protocol in terms of the framework above. We consider $\la \dots\ra_0$ as the Haar measure over the sphere $|\mathbf{m}|=1$: this corresponds, at finite size, to the infinite temperature state in the sector with maximal total spin $S$. In general, one can expect that the distribution spreads uniformly in time across the unit sphere, reaching eventually the infinite-temperature state ($\rho(t)\rightarrow 1$); specifically, the uniform distribution $\rho = 1$ is stationary, as easily checked. Nonetheless, other stationary states can appear, and this happens explicitly for the Ising model as shown in the next section.

\paragraph{Averaged dynamics and quantum trajectories---}

We study the evolution of symmetric distributions, corresponding (at finite-size) to diagonal density matrices in the basis of the Dicke states
\be\label{eq:Dicke}
\ket{N}\propto (S^+)^N\ket{\downarrow \dots \downarrow};
\ee
these are entirely characterized by the probability distribution of the magnetization density $m^z$. To obtain the drift and the diffusion of $m^z$, defined as its infinitesimal increase of mean and variance, we proceed as follows. We fix an initial state $\la\dots\ra$ with a given value of $m^z$ and we let it evolve infinitesimally via \eqref{eq:lind_classical}, obtaining (see End Matter)
\be\label{eq:semiclassical_ev}
\begin{cases}
\frac{d}{dt}\langle m^z \rangle = \frac{1}{2}\langle \{h,\{h,m^z\}\}\rangle\\
\frac{d}{dt}[\langle (m^z)^2 \rangle-\langle m^z \rangle^2] = \langle \{h,m^z\}^2\rangle.
\end{cases}
\ee
Denoting $m^z$ with $z$, and identifying the drift $\mu(z)$ and the diffusion $D(z)$, we express the evolution of the distribution $p(z)$ via the (Ito) Fokker-Planck equation
\be\label{eq:Fokker-Planck}
\partial_t p+\partial_z(\mu(z)p) = \frac{1}{2}\partial^2_z(D(z)p).
\ee

The quantum trajectories corresponding to this process can be identified. For instance, after each measurement outcome, the microscopic state is a Dicke state $\ket{N}$, whose thermodynamic limit is given by a probability distribution concentrated at $|\mathbf{m}|=1$ and with a given value of $m^z$ (see also Ref. \cite{Morettini-25}). Thus, we can associate an (Ito) Langevin equation with the dynamics of the magnetization as follows
\be\label{eq:Langevin}
dz = \mu(z)dt + \sqrt{D(z)}dW,
\ee
with $dW$ a Wiener process (whose time average is $\overline{dW^2} =dt$), playing the role of an unravelling of the averaged dynamics in Eq. \eqref{eq:Fokker-Planck}: this is represented in Fig. \ref{fig:sketch}. It is important to note that the emergence of a description in continuous time for the quantum trajectories, related to the fact that the measured observable $m^z$ becomes a continuous variable, is a property of the strict thermodynamic limit, and it does not hold at finite-size (in contrast to different protocols such as that of Ref. \cite{slcg-25}).

In the End Matter, we give the detailed calculation of $\mu(z)$ and $D(z)$ for the Ising model, reporting here the final result
\be\label{eq:mu_D}
\begin{cases}
\mu(z) = -\frac{z}{2}\l \lambda^2 + \frac{J^2}{2}(1-z^2)\r,\\
D(z) = \frac{J^2}{8}(1-z^2)^2+\frac{\lambda^2}{2}(1-z^2).
\end{cases}
\ee
Further, as we prove in the End Matter, the drift and the diffusion are related, as a consequence of the fact that $p(z)=1$ is a stationary state, and one can express \eqref{eq:Fokker-Planck} as
\be
-\partial_t p = -\frac{1}{2}\partial_z[D(z)\partial_z p].
\label{eq:markovoperator}
\ee
This formulation allows us to interpret the r.h.s. of the equation above as a (positive semidefinite) Markov operator \cite{Bakry-14} acting on the space of smooth functions of $[-1,1]$. From this point of view, it is immediate to check that $p(z)=1$ belongs to the discrete spectrum and it has a vanishing eigenvalue. Moreover, equation \eqref{eq:markovoperator} demonstrates that $p(z)=\delta(z \pm 1)$ are stationary states when $\lambda=0$, but not when $\lambda\neq 0$, as previously claimed. In the next section, we prove that, counterintuitively, this operator is gapped.

\paragraph{Eigenspectrum of the Markov operator---}

We study an operator of the form
\be\label{eq:Markov_op}
-\frac{1}{2}\partial_z[D(z)\partial_z \cdot ]
\ee
has a $L^2$ operator in of the interval $[-1,1]$ with respect to the flat (Lebesgue) measure $dz$. After a transformation, detailed in the End Matter, we obtain an equivalent Schr\"odinger operator of the form
\be\label{eq:Schr_op}
-\frac{1}{2}\partial^2_y  + V(y);
\ee
the change of variable is explicitly given by $dz = \sqrt{D(z)}dy$ and the potential $V(y)$ is\footnote{Here, with a slight abuse of notation, $D(y)$ denotes $D(z(y))$.}
\be
V(y) = \frac{1}{8}\frac{D''(y)}{D(y)}-\frac{3}{32}\l\frac{D'(y)}{D(y)}\r^2.
\ee
We focus on the critical point $\lambda=0$, choosing w.l.o.g. $J^2 = 8$ for simplicity, and we obtain
\be\label{eq:pot_ising}
z = \tanh(y), \quad V(y) = \tanh^2(y)-\frac{1}{2}.
\ee
Specifically, the interval $(-1,1)$ is mapped onto the infinite line $(-\infty,+\infty)$ and the potential satisfies $V(y\rightarrow \pm \infty) = 1/2$. This allows us to conclude that a discrete spectrum, associated with bound states, is separated from the continuous spectrum by a threshold at energy $E=1/2$. Moreover, analytical results for $V(y)$, being the P{\"o}schl-Teller potential at a reflectionless point, are known; the unique bound state with energy $E=0$, corresponding to the stationary state $p(z)=1$, is
\be
\psi(y) \propto \frac{1}{\cosh(y)},
\ee
while the continuous spectrum contains the wave functions
\be
\psi_k(y) \propto (\tanh(y)-ik)e^{iky},
\ee
with energy $E(k) = (k^2+1)/2$. In conclusion, the spectral gap above $E=0$ is finite, with gap $\Delta =1/2$.

We observe that the counterintuitive property of the operator~\eqref{eq:Markov_op}, defined on a finite interval, exhibiting a continuous spectrum arises primarily from the singular behaviour of $D(z)$ near $z=\pm 1$, where $D(z)\sim (z\mp 1)^2$.
 Namely, whenever $\lambda \neq 0$, then $D(z) \sim (z\mp 1)$ and the transformation $z\rightarrow y$ maps the interval $[-1,1]$ to another compact interval: in that case, the potential acts on a compact region and the spectrum is discrete (although we are not aware of general analytical results in this case).

So far, the analysis of the spectrum has been performed directly in the thermodynamic limit. For instance, it should be clear that while $p(z)\sim \delta(z\pm 1)$ are two stationary states, associated with the limit of the product states with spins aligned in the $z$ directions, they do not appear in the $L^2$-spectrum since they are singular distributions. In the next section, we analyse the finite-size Lindbladian, comparing its properties with the predictions from the Markov operator \eqref{eq:Markov_op}, showing compatibility in the low-lying spectrum.

\paragraph{Lattice results---}

To consider the spin chain at finite $L$ we express a symmetric mixed state $\rho$, supported in the sector with maximal spin, as follows
\be
\rho = \sum_N p_N \ket{N}\bra{N}
\ee
with the Dicke states defined in \eqref{eq:Dicke} and $\sum_Np_N = 1$. The time-evolution \eqref{eq:Lindblad} mixes the probabilities $p_N$, giving rise to an effective classical (continuous-time) Markov chain. We encode the evolution in a Markov matrix $W$ containing the transition rates for the process $N\rightarrow N'$: we do that using the Doi-Peliti formalism \cite{Doi-76,Doi-76a}, representing the state as
\be
|\rho)= \sum_N p_N|N),
\ee
with $\{|N)\}$ an orthonormal basis, and the evolution via $|\rho(t)) := e^{-Wt} |\rho)$. The matrix $W$ can be computed directly from Eq. \eqref{eq:Lindblad}, and, at the critical point $\lambda =0$, its non-vanishing matrix elements are (details in the End Matter)
\be\label{eq:Markov_Mat_el}
(N|W|N\pm 2) = -\frac{J^2}{16L^2}|\bra{N}(S^{\mp})^2\ket{N\pm 2}|^2;
\ee
the diagonal elements are fixed by the property $(\mathds{1}| W = 0$, with $(\mathds{1}|= \sum_N (N|$, which corresponds to conservation of probability.

At any finite-size $L$, the kernel of $W$ is generated by two stationary states
\be
|\rho) \propto \sum_{N\text{ even}}|N),\sum_{N\text{ odd}}|N).
\ee
This double degeneracy is related to the absence of hopping between Dicke states with different parity, a property that is lost for $\lambda \neq 0$ where the unique stationary state is $\rho \propto \sum_N |N)$. Crucially, these two states are locally indistinguishable $L\rightarrow \infty$, since they are both associated with the flat distribution $p(z) \propto 1$. In contrast, the two states $|N=0)$ and $|N=L)$ are never stationary at finite $L$, although they do not evolve in the thermodynamic limit since they correspond to the points $z\pm 1$ where the diffusion and the drift both vanish. Therefore, one could erroneously expect that such a mechanism is associated with the closure of the gap of $W$ at large $L$: however, this does not happen and the gap remains finite.

\begin{figure}[t]
 \includegraphics[width=\columnwidth]{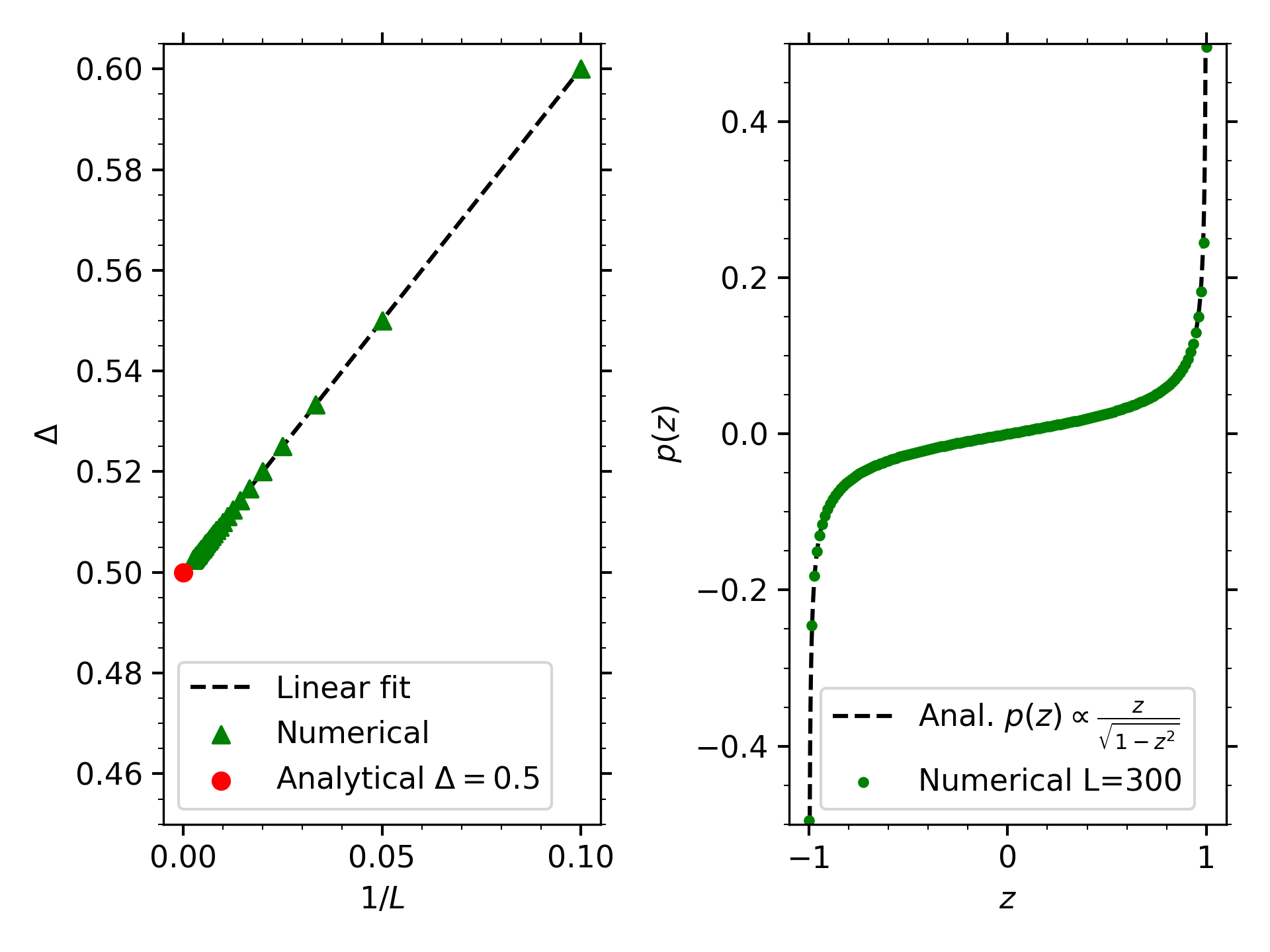}
 \caption{Left: Finite-size spectral gap $\Delta$ of the Markov operator $W$ in Eq. \eqref{eq:Markov_Mat_el} as a function of $1/L$, with $L\leq 400$. The data (green points) are consistent with a linear fit (black dashed line), and the extrapolation $\Delta \rightarrow 1/2$ for $L\rightarrow \infty$. Right: First eigenfunction above the gap computed at finite size ($L=300$), versus the corresponding prediction \eqref{eq:p_exc} for $L\rightarrow \infty$: an irrelevant proportionality constant, necessary to compare the two curves, has been fitted.}
\label{fig:spectrum}
\end{figure}

\begin{figure}[t]
 \includegraphics[width=\columnwidth]{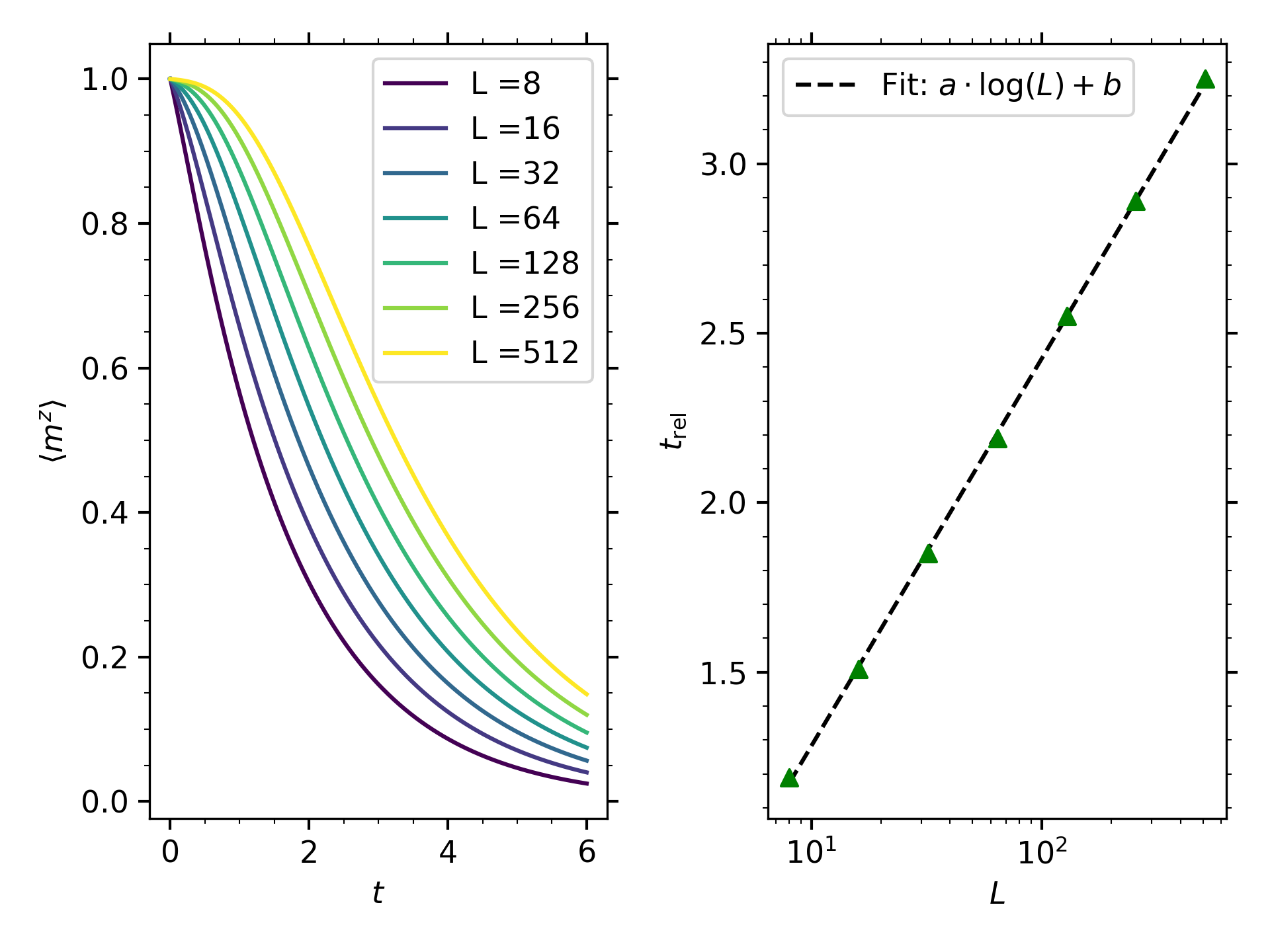}
 \caption{Left: Dynamics of the averaged magnetization $\langle m^z \rangle$ for different system sizes, starting from the fully polarized state. We observe a parametric slowdown as a function of $L$. Right: Relaxation time $t_{\text{rel}}$, identified by $\langle m^z \rangle=1/2$, as a function of $L$ (green markers). The data are consistent with a fit $t_{\text{rel}} = a \log L +b$ (black dashed line).}
\label{fig:relaxation}
\end{figure}

We choose $J^2 = 8$ and we plot in Fig. \ref{fig:spectrum} (left panel) the finite-size spectral gap $\Delta$ of $W$ as a function of $L$: the data are consistent with the limit $\Delta\rightarrow 1/2$ when $L\rightarrow \infty$, which is the expected value from the analytic analysis of the spectrum of Eq. \eqref{eq:Markov_op}. We also show (right panel) the finite-size eigenfunction and compare it with the corresponding analytical prediction for \eqref{eq:Markov_op}, namely
\be\label{eq:p_exc}
p(z) \propto \frac{z}{\sqrt{1 - z^2}};
\ee
we remark that the eigenfunction \eqref{eq:p_exc} belongs to the continuous spectrum \eqref{eq:p_exc}, and, consequently, it is non-normalizable and $\int^{1}_{-1} dz |p(z)|^2 = \infty$ holds. Finally, in Fig. \ref{fig:relaxation} we plot (left) the dynamics of the average magnetization $\la m^z\ra$ starting from the initial state $|N=L)$ for different sizes. The typical relaxation time grows with $L$; in the right panel, we show that the time $t_{\text{rel}}$ required to reach half of the initial value ($\langle m^z \rangle = 1/2$) is compatible with the scaling $t_{\text{rel}} \sim \log L$.

\paragraph{Conclusions---}
We have developed a general framework for the analysis of monitored mean-field models in the presence of projective measurements of extensive operators. In particular, we provided an explicit example in which the Lindbladian spectrum is gapped at a point exhibiting a phase transition, revealing a novel phenomenology that departs from standard expectations for DQPT and probe thereof.

Several important questions remain open. An immediate challenge is to determine whether the phenomenology we uncovered can persist in short-range systems, where the appearance of novel stationary states in the strict thermodynamic limit under bulk dissipation is far from understood. In such settings, it would be particularly interesting to investigate possible transitions between discrete and continuous spectra in the many-body Lindbladian. Another direction is the development of efficient numerical strategies capable of extracting the structure of stationary states of infinite systems from the finite-size spectrum, which may in turn motivate sharper diagnostic tools for DQPT.

Finally, while our analysis focused primarily on situations where $[H,Q]\neq 0$, one may instead consider regimes in which a charge is globally conserved but measured on a subsystem (a setup which is closely related to recent studies on entanglement asymmetry \cite{Ares2023}), or weakly broken, potentially leading to emergent slow dynamics of Gibbs states (analogous to Ref. \cite{Lumia-25}).

\paragraph{Acknowledgements ---}

We thank L. Piroli and M. Mazzoni for sharing their preliminary results on related subjects and their collaboration in the early stage of the project. We are also grateful to A.Delmonte, G.Chiriac\`o, and G.Giachetti for interesting discussions. LC acknowledges support from the ANR project LOQUST ANR-23-CE47-0006-02. RT acknowledges support by the ERC-AdG grant MOSE No. 101199196.

\begin{center}
\begin{large}
\textbf{End Matter}
\end{large}
\end{center}

\paragraph{Drift and diffusion for the Ising model---}

In this section, we first derive Eq. \eqref{eq:semiclassical_ev} and sketch the explicit computation for the Ising model \eqref{eq:mu_D}. As a preliminary lemma, it is worth studying the evolution of $\la\mathcal{O}\ra$ under \eqref{eq:lind_classical} for a generic observable $\mathcal{O}$ that is symmetric under $z$-rotations (as for $m^z$ and $(m^z)^2$). Using that the corresponding Lindbladian is hermitian, it is possible to write
\be\label{eq:evol_O_sym}
\begin{split}
\frac{d}{dt}\la \mathcal{O}\ra = &\frac{1}{2}\int^{2\pi}_0 \frac{d\alpha}{2\pi}\langle\{h_\alpha,\{h_\alpha,\mathcal{O}\}\}\rangle = \\
&\frac{1}{2}\int^{2\pi}_0 \frac{d\alpha}{2\pi}\langle(\{h,\{h,\mathcal{O}\}\})_\alpha\rangle
\end{split}
\ee
where we have used $\mathcal{O}_\alpha = \mathcal{O}$, since the observable is symmetric, and the compatibility between Poisson brackets and rotations. If we further consider symmetric states, as in our protocol, it is possible to get rid of the integral over $\alpha$ in Eq. \eqref{eq:evol_O_sym}: this gives directly the equation for the evolution of $\la m^z\ra$ in Eq. \eqref{eq:semiclassical_ev}. We study the infinitesimal increase of variance for a state $\langle \dots\rangle$ that has a definite value of $m^z$, implying that $m^z$ is not correlated with other observables (meaning $\la m^z\dots\ra = \la m^z\ra \la\dots\ra $): this allows to simplify the following expression
\be\begin{split}
\frac{d}{dt}[\la (m^z)^2\ra -\la (m^z)\ra^2] = \frac{1}{2}\la \{h,\{h,(m^z)^2\}\}\ra - \\
\la m^z\ra\la \{h,\{h,m^z\}\}\ra
\end{split}
\ee
obtaining that of Eq. \eqref{eq:semiclassical_ev}.

For the Ising model, we focus on the thermodynamic limit of Dicke states, satisfying $\mathbf{m}^2=1$ and $m^z = \la m^z\ra$ inside the correlators, and we obtain
\be\begin{split}
&\frac{d}{dt}\la m^z\ra = \\
&\frac{J^2}{8}\la (m^{x})^2\{m^x,\{m^x,m^z\}\}\ra + \frac{\lambda^2}{2}\la\{m^x,\{m^x,m^z\}\}\ra
\end{split}
\ee
giving directly, after simple calculations, Eq. \eqref{eq:mu_D}. Similar steps can be performed to compute the diffusion; alternatively, one could employ the relation Eq. \eqref{eq:mu_D_relation}, together with the property $D(z=\pm 1) =0$ (required for consistency, being the dynamics restricted to $z \in [-1,1]$), obtaining the same result.

\paragraph{General relation between drift and diffusion---}

In this section, we prove the relation
\be\label{eq:mu_D_relation}
\mu(z) = \frac{1}{2}D'(z)
\ee
entering Eq. \eqref{eq:Markov_op}. Since the Lindbladian \eqref{eq:Lindblad} is hermitian, the Markov operator
\be
-\partial_z[\mu(z)\cdot] + \frac{1}{2}\partial_z^2[D(z)\cdot]
\ee
associated with the Fokker-Planck in Eq. \eqref{eq:Fokker-Planck}, has to be hermitian as well (w.r.t. the $L^2$-product associated with the measure $dz$ in $[-1,1]$). We rewrite as
\be\label{eq:split_markov}
\partial_z[(-\mu +\frac{1}{2}D')\cdot] + \frac{1}{2}\partial_z[D\partial_z \cdot],
\ee
observing that, since $\partial_z$ is antihermitian, then the second term in \eqref{eq:split_markov} is hermitian. Therefore, we also require that the first term is hermitian: after defining $B(z):= -\mu+D'/2$, we equivalently express the condition above as
\be\label{eq:cond_Bz}
\{\frac{d}{dz},B(z)\} =0,
\ee
with $\{\cdot,\cdot\}$ the anticommutator. The relation \eqref{eq:cond_Bz} has to be valid as an operator, that is, the l.h.s. has to annihilate every smooth function over $[-1,1]$: one can easily show that this stringent condition implies $B(z)=0$, and therefore the first term in Eq. \eqref{eq:split_markov} vanishes.

We remark on a technical, though subtle, point. The anti-Hermitian nature of $\partial_z$ can be verified on a suitable dense set of smooth functions with prescribed boundary conditions. In this context, it is natural to impose Neumann boundary conditions (e.g.\ $p'(-1)=p'(1)=0$), since they are satisfied by the stationary state $p(z)\propto 1$. On the other hand, it is generally not true that the eigenfunctions obey these boundary conditions; this is especially evident in our case, where a continuous spectrum appears and non-normalizable eigenfunctions—diverging at the boundary points—are present (see Eq.~\eqref{eq:p_exc}). We stress, finally, that different boundary conditions define, in principle, different self-adjoint extensions of a given differential operator \cite{Bakry-14}, and hence different spectra. This is the case for the Laplacian on $[-1,1]$, for which, for example, Neumann or Dirichlet ($p(-1)=p(1)=0$) boundary conditions may be considered. Here, a potential issue is present, since $D(z)$ vanishes at the boundary point: luckily, a suitable transformation, discussed in the next section, remedies this singular behaviour.

\paragraph{From the Fokker-Planck to the Schr\"odinger equation---}

Here, we explain the mapping from the Markov operator \eqref{eq:Markov_op} to \eqref{eq:Schr_op}. We first apply the change of variables $dz = \sqrt{D}dy$, representing the operator as
\be
-\frac{1}{2}\partial_y^2 - \frac{1}{4}\frac{\partial_yD}{D}\partial_y.
\ee
This allows us to get rid of the inhomogeneity in the second derivative term. In these coordinates, the $L^2$-product becomes
\be
\la p_0,p_1\ra := \int dz \overline{p_0(z)}p_1(z) = \int dy \sqrt{D} \ \overline{p_0(y)}p_1(y),
\ee
meaning that the (Lebesgue) measure $dz$ is mapped onto a new measure that, in general, is not flat. At this point, it is natural to perform a local rescaling of the function to overcome this issue, defining
\be
\psi(y) := D^{1/4}p(y),
\ee
and representing the operator in terms of its action on $\psi(y)$. This is straightforward, although lengthy, and the final result is \eqref{eq:Schr_op}. We finally observe that the absence of a first derivative term in \eqref{eq:Schr_op} is not surprising, since the operator is hermitian w.r.t. the $L^2$-product associated with the flat measure $dy$.

\paragraph{Matrix elements of the Markov operator---}

Starting from the the Lindbladian \eqref{eq:Lindblad}, and using the explicit expression for the Ising Hamiltonian at the critical point ($\lambda=0$), it is not hard to identify Lindblad operators (dissipators) $\propto (S^+)^2,(S^-)^2$, obtaining, after straightforward calculations, the matrix elements of $W$ in Eq.~\eqref{eq:Markov_Mat_el}. A closed formula can be provided, since the matrix elements of $(S^\pm)^2$ in a generic spin-$S$ representation are known analytically (see Ref.~\cite{Sakurai-17}), and for $J^2=8$ we have
\be
(N|W|N+2) = -\frac{N(N-1)(L+2-N)(L+1-N)}{2L^2}.
\ee

We observe that, in the (\textit{boundary}) limit where $L\rightarrow \infty$ with $N$ (or $L-N$) fixed, the rate converges to the finite value
\be
(N|W|N+2) \simeq -\frac{N(N-1)}{2}.
\ee
Therefore, it is misleading to think that the state $|N=0),|N=L)$ effectively decouples in the thermodynamic limit, even though the drift appearing in the Fokker-Planck equation vanishes at $z=\pm 1$. In contrast, if one performs the (\textit{bulk}) limit with $N/L = (1+z)/2$ fixed as $L\rightarrow \infty$, one obtains
\be
(N|W|N+2) \simeq -\frac{L^2}{2}\left( \frac{1-z^2}{4} \right)^2.
\ee
The diverging factor $\sim L^2$ in front is not surprising, and its presence originates from a lattice discretization of the Laplacian with lattice spacing $\propto 1/L$.

\bibliography{bibliography}

\onecolumngrid
\break
\begin{center}
    {\large \bf Supplemental Material to: \\
            Phase Transitions without gap closing in monitored quantum mean-field systems
}

\vspace{0.75cm}
\end{center}

\section{Imperfect measurements}

In this section, we comment on a slightly more realistic scenario, in which at each time step the measurement is performed with some probability $p$, as it would happen for an imperfect measurement apparatus. If one applies the unitary $U:= e^{-iH\tau}$ first and then the measurement, the evolution step of the density matrix now reads as
\begin{equation}
    \rho \mapsto (1-p)U\rho U^\dagger + p \sum_q \Pi_q U \rho U^\dagger \Pi_q.
\end{equation}
 Choosing $p=\gamma \tau$ and performing the small $\tau$ limit with $\tau= dt$, one obtains
\be\label{eq:Lind_imp}
\frac{d\rho}{dt} = -i[H,\rho]-\gamma \left(\rho- \sum_q \Pi_q \rho \Pi_q\right).
\ee 

Before entering the details, we comment on some general features of Eq. \eqref{eq:Lind_imp}. The dissipative term tends to symmetrize the state, while the unitary dynamics, driven by $H$, do not share this property: this is slightly different with respect to Eq. \eqref{eq:Lindblad}, where the state is symmetric at any time $t$. Also, the infinite temperature state is stationary, as one can check directly, and, in the absence of strong symmetries (say, operators $\mathcal{O}$ commuting with $H$ and $Q$), this is expected to be the only one (at least at finite size) \cite{bp-12,aj-14}. Further, when the Hamiltonian $H$ vanishes, the solution of \eqref{eq:Lind_imp} becomes simple. For instance, given $p_t := e^{-\gamma t}$, we easily compute
\be\label{eq:lind_dissip}
\rho(t) = p_t \rho(0) + (1-p_t) \sum_q \Pi_q \rho(0) \Pi_q,
\ee
which is a mixture of the initial state and the corresponding symmetrisation; in particular, the state becomes symmetric after a typical time $\sim \gamma^{-1}$. Finally, it is worth observing that, in general, the dynamics implied by \eqref{eq:Lind_imp} is non-local when the charge $Q$ is extensive: this is the case both for many-body systems with local hamiltonians, due to the non-locality of the projectors $\Pi_q$, and for the mean-field models considered in the main-text, since the symmetrisation is non-local in the semiclassical phase-space (in contrast with Eq. \eqref{eq:lind_classical}, which gives rise to a local Fokker-Planck equation). Nonetheless, since this issue comes from the symmetrisation only, one can expect that, in the limit of large $\gamma$, the evolution becomes similar to that of the main text. We show that explicitly, using second-order perturbation theory, writing down the corresponding effective Lindbladian (see Ref. \cite{Cai-13,Marche-25} for details). 

We first split the Lindbladian \eqref{eq:Lind_imp} as follows
\begin{equation}
    \mathcal{L}[\rho] = \mathcal{L}_1 [\rho] + \gamma  \mathcal{L}_2 [\rho],
\end{equation}
and then, we identify the two eigenspaces of $\mathcal{L}_2$: one is generated by the symmetric states, whose corresponding eigenvalue is $0$, and its orthogonal complement has eigenvalue $-\gamma$. Then, after simple calculations, we write down the effective Lindbladian associated with the kernel of $\mathcal{L}_2$ as
\begin{equation}
    \mathcal{L}_{\rm eff}[\rho] = -i\int\frac{d\alpha}{2\pi}e^{i\alpha Q}[H,\rho]e^{-i\alpha Q}  -\frac{1}{\gamma}  \int \frac{d\alpha}{2\pi} e^{i\alpha Q} [H,[H,\rho]]e^{-i\alpha Q},
\end{equation}
acting on the space of symmetric states. The first term is just the symmetrized unitary evolution and, in the context of the mean-field dynamics restricted to the space of Dicke states, it vanishes as stressed in the main text. The second one, instead, is precisely \eqref{eq:Lindblad} up to a time redefinition.

\end{document}